\documentstyle[aps,12pt,manuscript]{revtex}
\begin{document}
\preprint{INJE--TP--95--6 }
\def\overlay#1#2{\setbox0=\hbox{#1}\setbox1=\hbox to \wd0{\hss #2\hss}#1%
\hskip -2\wd0\copy1}

\title{On the stability of two-dimensional extremal black holes}

\author{  Y. S. Myung }
\address{Department of Physics, Inje University, Kimhae 621-749, Korea}

\maketitle
\vskip 1.5in

\begin{abstract}
We discuss the stability of the  extremal ground states of  a
two-dimensional (2D)
 charged black hole which carries  both electric ($Q_E$) and magnetic
($Q_M$) charges.
The method is first
to find the physical field and then to derive the equation of the
Schr\"odinger type.
It is found that the presenting potential to an on-coming tachyon (as a
spectator) takes a
 barrier-well type. This provides the  bound state solution, which
implies  an exponentially growing mode with respect to time.
The 2D extremal ground states all  are
classically unstable. We conclude that the 2D  extremal charged black holes are
not considered as the candidates for the stable endpoint of the Hawking
evaporation.

\end{abstract}

\newpage
Recently the extremal black holes have received much attention.  Extremal
black holes provide
a simple laboratory in which to investigate the quantum aspects of black
hole [1]. One of the crucial
features is that the Hawking temperature vanishes.  The black hole with $M>Q$
  will tend to Hawking
radiate down to its extremal $M=Q$ state. Thus the extremal black hole may
play  a role of
the stable endpoint for the Hawking evaporation. It has been also proposed
that although
the extremal black hole has nonzero area, it has zero entropy [2]. This is
because the extremal case is
distinct topologically  from the nonextremal one.

It is very important to enquire into the stability of the extremal black
holes, which establishes their
physical existence. It has been shown that the 4D extremal  charged black
holes including
the Reissner-Nordstr\"om one are
shown to be classically  stable [3]. Since all potentials are positive
definite outside the
outer horizon, one can infer  the stability of 4D extremal  charged black
holes  using the same
argument as employed by Chandrasekhar [4]. However,
it was demonstrated that the 2D electrically  charged extremal black hole
is unstable [5].

In this letter, we shall perform a complete analysis of the stability for
the 2D  extremal black holes
with electric and magnetic charges. One easy way of understanding  a black
hole is to find
out how it reacts to external perturbations.
We always  visualize the
black hole as presenting an effective potential barrier (or well) to the
on-coming waves [4].
In deciding whether or not the the extremal black hole is stable, one
starts with a physical
perturbation which is regular everywhere in space at the initial time $t=0$
[6]. And then see whether such a
perturbation will grow with time. If there exists an exponentially growing
mode, the extremal
black hole is unstable. As a  compact criterion, the extremal  black bole
is unstable if it has
the potential well to the on-coming waves [7]. This is so because in the
Schr\"odinger equation the potential well always allows
the bound state as well as scattering states. The former shows up as an
imaginary frequency mode,
leading to an exponentially growing mode.

We  start with  the low energy action from heterotic string theory [8]
\begin{equation}
S_{l-e} = \int d^2 x \sqrt{-G} e^{-2\Phi}
   \big \{ R + 4 (\nabla \Phi)^2  - {1 \over 2} F^2 - {1 \over 2} (\nabla
T)^2 + V(\Phi,T)\big \}
\end{equation}
with the potential $V(\Phi,T) =\alpha^2 -Q_M^2 e^{4\Phi} +T^2.$
Here are all string fields ( metric $G_{\mu\nu}$,  dilaton $\Phi$,  Maxwell
field $F_{\mu\nu}$,
 and tachyon $T$). We introduce the tachyon as a spectator and the term
$-Q_M^2 e^{4\Phi}$
for a magnetically charged configuration. From the  4D magnetic Mawxell
field ($F= Q_M \sin \theta d
\theta \land d\varphi)$, one finds the modification in two dimensions :
$V(0,T) \to V(\Phi,T)$. Also this type of dilaton
potential can be generated from the closed string loop corrections.
Setting $\alpha^2 = 8$ and  after deriving equations, we take the
transformation
\begin{equation}
-2\Phi \rightarrow \Phi,~~~ T \rightarrow \sqrt 2 T, ~~~-R  \rightarrow R.
\end{equation}
Then the equations of motion become
\begin{eqnarray}
&&R_{\mu\nu} + \nabla_\mu \nabla_\nu \Phi  + \nabla_\mu T \nabla_\nu T +
F_{\mu\rho}F_{\nu}^{~\rho}
 - Q^2_M e^{-2 \Phi} G_{\mu\nu}= 0,  \\
&& \nabla^2 \Phi + (\nabla \Phi)^2  - {1 \over 2} F^2 + Q^2_M e^{-2 \Phi} -
2 T^2 - 8 = 0,  \\
&&\nabla_\mu F^{\mu \nu} + (\nabla_\mu \Phi) F^{\mu \nu} = 0,   \\
&&\nabla^2 T + \nabla \Phi \nabla T + 2 T = 0.
\end{eqnarray}

An electrically  and magnetically charged black hole solution to the above
equations is given by

\begin{equation}
\bar \Phi = 2 \sqrt 2 r,~~~ \bar F_{tr} = Q_E e^{-2 \sqrt 2 r},~~~ \bar T = 0,
{}~~~ \bar G_{\mu\nu} =
 \left(  \begin{array}{cc} - f & 0  \\
                            0 & f^{-1}   \end{array}   \right),
\end{equation}
with
\begin{equation}
f = 1 -  {2M \over 2\sqrt 2}e^{- 2 \sqrt 2 r} + {Q^2 \over 8}e^{- 4 \sqrt 2 r},
\end{equation}
where $M$ and $Q = \sqrt{Q^2_E +Q^2_M}$ are the mass and total charge of
the black hole, respectively.
Here we take $ M=\sqrt2$ for convenience.    For $0< Q <M$, the double
horizons ($r_{\pm}$) are given by
\begin{equation}
r_{\pm} = { 1 \over 2 \sqrt 2 }\log \left[ {1 \pm  \sqrt { 1 - {Q^2 \over
2}} \over 2}\right],
\end{equation}
where $r_{+}(r_{-})$ correspond to the event (Cauchy) horizons.
This charged black hole may provide an ideal setting for studying the late
stages
of Hawking evaporation. For $Q=M$, two horizons coincide :$
r_{+}=r_{-}\equiv r_o$.
We are mainly interested  in this extremal limit.

To study the propagation of string fields, we introduce small perturbation
fields  around
the background solution as [5]
\begin{eqnarray}
&&F_{tr} = \bar F_{tr} + {\cal F}_{tr} = \bar F_{tr} [1 - {{\cal F}(r,t)
\over Q_E}],        \\
&&\Phi = \bar \Phi + \phi(r,t),                       \\
&&G_{\mu\nu} = \bar G_{\mu\nu} + h_{\mu\nu}  = \bar G_{\mu\nu} [1 - h
(r,t)],     \\
&&T = \bar T + \tilde t \equiv \exp (-{\bar \Phi \over 2}) [ 0 + t (r,t) ].
\end{eqnarray}
 One has to linearize (3)-(6) in order to obtain the equations governing
the perturbations as
\begin{equation}
  \delta R_{\mu\nu} (h)
+ \bar \nabla_\mu \bar \nabla_\nu \phi
- \delta \Gamma^\rho_{\mu\nu} (h) \bar \nabla_\rho \bar \Phi
+ 2 \bar F_{\mu \rho} {\cal F}_\nu^{~ \rho}
- \bar F_{\mu \rho} \bar F_{\nu\alpha} h^{\rho \alpha}
+Q^2_M e^{- 4\sqrt 2 r} \bar G_{\mu\nu}(h+ 2\phi)
 = 0,
\end{equation}

\begin{eqnarray}
&&  \bar \nabla^2 \phi
- h^{\mu\nu} \bar \nabla_\mu \bar \nabla_\nu \bar \Phi
- \bar G^{\mu\nu} \delta \Gamma^\rho_{\mu\nu} (h) \partial_\rho \bar \Phi
- h^{\mu\nu} \partial_\mu \bar \Phi \partial_\nu \bar \Phi
+ 2 \bar G^{\mu\nu} \partial_\mu \bar \Phi \partial_\nu \phi
\nonumber   \\
&&~~~~~~~~~~~~~~~- \bar F_{\mu \nu} {\cal F}^{\mu \nu}
     + \bar F_{\mu \nu}   \bar F_{\rho}^{~\nu} h^{\mu\rho} - 2 Q^2_M e^{-
4\sqrt 2 r} \phi = 0,
\end{eqnarray}

\begin{equation}
  ( \bar \nabla_\mu + \partial_\mu \bar \Phi )
    ( {\cal F}^{\mu \nu} - \bar F_\alpha^{~~\nu} h^{\alpha \mu}
                             - \bar F^{\mu}_{~~\beta} h^{\beta \nu} )
      + \bar F^{\mu \nu} (\delta \Gamma^\sigma_{\sigma \mu} (h)
      + (\partial_\mu \phi))
 = 0,
\end{equation}

\begin{equation}
\bar \nabla^2 \tilde t + \bar \nabla_\mu \bar \Phi \bar \nabla^\mu \tilde t
+ 2 \tilde t = 0,
\end{equation}
where

\begin{eqnarray}
&&\delta R_{\mu\nu} (h) = {1 \over 2} \bar \nabla_\mu \bar \nabla_\nu
h^\rho_{~\rho}
 + {1 \over 2} \bar \nabla^\rho \bar \nabla_\rho h_{\mu\nu}
 - {1 \over 2} \bar \nabla^\rho \bar \nabla_\nu h_{\rho\mu}
 - {1 \over 2} \bar \nabla^\rho \bar \nabla_\mu h_{\nu\rho},   \\
&&\delta \Gamma^\rho_{\mu\nu} (h) = {1 \over 2} \bar G^{\rho\sigma}
( \bar \nabla_\nu h_{\mu\sigma} + \bar \nabla_\mu h_{\nu\sigma} - \bar
\nabla_\sigma h_{\mu\nu} ).
\end{eqnarray}

 From (16) one can express ${\cal F}$ in terms of $\phi$ and $h$ as

\begin{equation}
{\cal F} = - Q_E ( \phi + h ).
\end{equation}
This means that ${\cal F}$ is no longer an independent mode.
Also from the diagonal element of (14), we have

\begin{eqnarray}
&&  \bar \nabla^2 h - 2 \bar \nabla^2_t \phi - 2 \sqrt 2 \bar G^{rr}
\partial_r h
       - 2 Q^2 e^{- 4 \sqrt 2 r} (h + 2 \phi) = 0,    \\
&&  \bar \nabla^2 h - 2 \bar \nabla^2_r \phi + 2 \sqrt 2 \bar G^{rr}
\partial_r h
       - 2 Q^2 e^{- 4 \sqrt 2 r}(h + 2 \phi) = 0.
\end{eqnarray}
Adding the above two equations leads to

\begin{equation}
\bar \nabla^2 ( h - \phi) -
   2 Q^2 e^{- 4 \sqrt 2 r} (h + 2 \phi) = 0.
\end{equation}
Also  the dilaton equation (15)  leads to
\begin{equation}
\bar \nabla^2 \phi + 4 \sqrt 2 f \partial_r \phi + 2 \sqrt 2 (\partial_r f
+ 2 \sqrt2 f) h
- 2 Q^2 e^{- 4 \sqrt 2 r}\phi = 0.
\end{equation}
And the off-diagonal element of (14) takes the form
\begin{equation}
 \partial_t \big\{ ( \partial_r  - \Gamma^t_{tr}) \phi + \sqrt 2  h  \big\}
   = 0,
\end{equation}
which provides us  the relation between $\phi$ and $h$ as
\begin{equation}
\partial_r \phi = -\sqrt2 h + {1 \over 2}{\partial_r f \over f} \phi + U(r).
\end{equation}
Here $U(r)$ is the residual gauge degrees of freedom and thus we set $U(r)
= 0$ for simplicity.
Substituting (26) into (24), we have
\begin{equation}
\bar \nabla^2 \phi + 2 \sqrt 2  \partial_r f ( h + \phi)
- 2 Q^2 e^{- 4 \sqrt 2 r} \phi = 0.
\end{equation}
Calculating (23) + 2 $\times$ (27), one finds the other equation
\begin{equation}
\bar \nabla^2 (h + \phi) + 4 \sqrt 2  \partial_r f ( h + \phi)
- 2 Q^2 e^{- 4 \sqrt 2 r} ( h + 4 \phi) = 0.
\end{equation}
Although (23) and (27) look like the very complicated forms,
these reduce to
\begin{eqnarray}
&&\bar \nabla^2 ( h - \phi) = 0, \\
&&\bar \nabla^2 (h + \phi) + 4 \sqrt 2  \partial_r f ( h + \phi) =0
\end{eqnarray}
in the asymptotically flat region ($r \to \infty$).
This suggests that one  obtain  two graviton-dilaton modes.
However, it is important to check whether the graviton ($h$),  dilaton
($\phi$), Maxwell
mode (${\cal F}$)  and tachyon ($t$) are  physically propagating modes
in the 2D charged black hole background.
We review the conventional counting of degrees of freedom.
The number of degrees of freedom for the gravitational field ($h_{\mu\nu}$) in
$D$-dimensions is $(1/2) D (D -3)$.  For a Schwarzschild black hole,
we obtain two degrees of freedom. These correspond to the Regge-Wheeler
mode for odd-parity perturbation
and Zerilli mode for even-parity perturbation [4].  We have $-1$ for $D=2$.
This means that in
two dimensions
the contribution of the graviton is equal and opposite to that of a
spinless particle (dilaton).
The graviton-dilaton modes ($h+\phi, h-\phi$) are gauge degrees of freedom
and thus turn out to be
nonpropagating modes[6].
In addition, the Maxwell field has $D-2$ physical degrees of freedom.
The Maxwell field has no physical degrees of freedom for $D=2$. Actually
from (20) it turns out to be
a redundant one.
Since these all are  nonpropagating modes, it is  necessary to consider the
remaining one (17).
The tachyon as a spectator is a physically propagating mode.
This is used to illustrate many of the qualitative results about the 2D
charged black hole
in a simpler context.
The stability should be based on the physical degrees of freedom.
Its linearized equation is
\begin{equation}
f^2 t'' + ff't' -[\sqrt 2 f f' -2 f(1-f)]t  - \partial_t^2 t = 0,
\end{equation}
where the prime ($\prime$) denotes the derivative with respect to $r$.
To study the stability,  the above equation
should be transformed into  one-dimensional Schr\"odinger equation.
Introducing a tortoise coordinate
$$r\to r^* \equiv g(r),$$
(31) can be rewritten as
\begin{equation}
f^2 g'^2 {\partial^2 \over \partial r^{*2}} t  + f \{ f g'' +  f' g'\}
{\partial \over \partial r^* }t - [\sqrt 2 ff' - 2 f (1 - f)]t
 - {\partial^2 \over \partial t^2} t = 0.
\end{equation}
Requiring that the coefficient of the linear derivative vanish, one finds
the relation
\begin{equation}
g' =  {1 \over f}.
\end{equation}
Assuming $t( r^*,t ) \sim \tilde t ( r^* ) e^{i\omega t}$,
one can cast (32) into the Schr\"odinger equation

\begin{equation}
\{ {d^2 \over dr^{*2}} + \omega^2 - V(r)\} \tilde t = 0,
\end{equation}
where the effective potential $V(r)$  is given by
\begin{equation}
V(r) = f(\sqrt 2 f' - 2  (1 - f)).
\end{equation}
 Fig. 1 shows  the graphs of potentials for $Q=0.1, 1,$ and $\sqrt 2$. When
$M(=\sqrt 2)>Q$,
the potentials outside the
event horizon are simple barriers. However  a  barrier-well potential
appears outside the
event horizon when the nonextremal
black hole (a simple barrier) approaches the extremal one.
This takes the form
\begin{equation}
V_{extr}(r)  = 2 e^{- 2 \sqrt 2 r} ( 1- {1 \over 2}e^{- 2 \sqrt 2 r})^2 (
1- {3 \over 4}e^{- 2 \sqrt 2 r}) .
\end{equation}
The event horizon is located at $ r_o= -0.245$.
Now let us translate the potential $V_{extr}(r)$ into $V_{extr}(r^*)$.
With $f= (1- {1 \over 2} \exp(-2\sqrt 2 r))^2$, one obtains the explicit
form of $r^*$
\begin{equation}
r^*= r - {1 \over 2\sqrt 2 (1- {1 \over 2} e^{-2\sqrt 2 r})} +
    {1 \over 2 \sqrt 2} \log |1- {1 \over 2} e^{-2\sqrt 2 r}|.
\end{equation}
Since both the forms of $V_{extr}(r)$ and $r^*$ are very complicated, we
are far from obtaining
the exact form of $V_{extr}(r^*)$. Instead we can find an approximate form.
{}From (37), in the asymptotically flat region one finds that $r^* \simeq r$.
(36) takes
the asymptotic form
\begin{equation}
V_{r*\to \infty} \simeq 2 \exp( -2 \sqrt 2 r^*).
\end{equation}
On the other hand, near the horizon ($r=r_o$) one has
\begin{equation}
r^* \simeq - { 1 \over 2 \sqrt 2( 1 - {1 \over 2} e^{ - 2 \sqrt 2 r})}.
\end{equation}
Approaching the horizon $(r\to r_o, r^* \to -\infty)$, the potential takes
the form
\begin{equation}
V_{r*\to -\infty} \simeq - {1 \over 4 r^{*2}}.
\end{equation}
Using (38) and (40) one can construct the approximate form $V_{app}(r^*)$
(Fig. 2).
This is also a  barrier-well which is  localized at the origin of $r^*$.
Our stability analysis is  based on the equation
\begin{equation}
\{ {d^2 \over dr^{*2}} + \omega^2 - V_{app}(r^*)\} \tilde t = 0.
\end{equation}

As is well known, two kinds of solutions to the Schr\"odinger equation
correspond to
 the bound and scattering states. In our case $V_{app}(r^*)$ admits  two
solutions  depending on
the signs of the energy.

(i) For $E>0 (\omega=$ real), the asymptotic solution for $\tilde t$ is given
by

\begin{eqnarray}
\tilde t_\infty & = & \exp(i \omega r^*)  + R \exp(- i \omega r^*)~~~~~~~~
( r^*  \to \infty ),  \\
\tilde t_{EH} & = & T\exp( i\omega r^*)~~~~~~~~~~~~~~~~~~~~~~~~~~~
( r^*  \to - \infty ),
\end{eqnarray}
where $R$ and $T$ are the scattering amplitudes of two waves which are
reflected and transmitted by the potential $V_{app}(r^*)$, when a tachyonic
wave of unit
amplitude with the frequency $\omega$ is incident on the black hole from
infinity.

\noindent $(ii)$ For $E<0 (\omega =-i \alpha$, $\alpha$ is positive and real),
 we have the bound state.
Eq. (41) and possible asymptotic solutions are given by

\begin{eqnarray}
{d^2 \over d r^{*2}}\tilde t & = & (\alpha^2 + V_{app}(r^*)) \tilde t,     \\
\tilde t_\infty & \sim & \exp(\pm \alpha r^*),~~~~~~~~ ( r^*  \to \infty )  \\
\tilde t_{EH}   & \sim & \exp(\pm \alpha r^*) ~~~~~~~ ( r^*  \to - \infty ).
\end{eqnarray}
To ensure that the perturbation falls off to zero for large $r^*$, we choose
$\tilde t_\infty \sim \exp (-\alpha r^*)$.  In the case of $\tilde t_{EH}$,
the solution
$\exp (\alpha r^*)$ goes to zero as $r^* \to - \infty$.
Now let us observe whether or not $\tilde t_{EH} \sim \exp (\alpha r^*)$
can be matched
to $\tilde t_\infty \sim \exp (-\alpha r^*)$.
Assuming $\tilde t$ to be positive,
the sign of $d^2 \tilde t / dr^{*2}$
can be changed from $+$ to $-$ as  $r^*$
goes from $\infty$ to $-\infty$.
If we are to connect $\tilde t_{EH}$ at one end to a decreasing solution
$\tilde t_\infty$
at the other, there must be a point ($d^2\tilde t/ dr^{*2}<0$,
$d \tilde t/dr^*=0$) at which the signs of $\tilde t$ and
$d^2\tilde t/dr^{*2}$ are opposite : this is  compatible with  the shape of
$V_{app}(r^*)$ in Fig. 2. It thus is possible for
$\tilde t_{EH}$ to be connected to $\tilde t_\infty$ smoothly.  Therefore a
bound state solution
is given by

\begin{eqnarray}
\tilde t_\infty & \sim & \exp(- \alpha r^*),~~~~~~~~ ( r^*  \to \infty )     \\
\tilde t_{EH}   & \sim & \exp( \alpha r^*) ~~~~~~~~ ( r^*  \to - \infty ).
\end{eqnarray}
This is a regular solution everywhere in space at the initial time $t=0$.
It confirms this solution from the quantum machanics which states that  the
bound state
solution is always allowed if a potential well exists.
However, on later $\omega=-i\alpha$ implies
$t_\infty (r^*, t)=\tilde t_\infty(r^*)\exp(-i\omega t) \sim \exp (-\alpha
r^*) \exp (\alpha t)$ and
$t_{EH} (r^*,t)=\tilde t_{EH}(r^*) \exp(-i\omega t) \sim \exp (\alpha r^*)
\exp (\alpha t)$.
This means
that there exists an exponentially growing mode with time.
Therefore, the 2D  extremal ground state  is
classically unstable. The origin of this instability comes from  a
barrier-well.
This potential appears when the nonextremal black hole approaches the
extremal limit.
As is discussed in Ref. [9], the quantum stress tensor of a scalar field
(instead of the tachyon)
in the extremal black hole
diverges at the horizon. This means that the 2D extremal black hole is also
quantum-mechanically unstable.
This divergence can be better understood by the regarding an extremal black
hole as the limit of a
nonextremal one. A nonextremal black hole has an outer (event) and an inner
(Cauchy) horizon,
 and these come together
in the extremal limit.  In this case, we find that if we adjust  the
quantum state of the scalar
 field so that  the stress tensor is finite at the outer horizon, it always
diverges at
the inner horizon.
Thus it is not so surprising that in the extremal limit (when the two
horizons come together)
the divergence persists, although it has a softened form.
By the similar way, it conjectures that the classical instability
originates from the instability
(blueshift) of the inner horizon.
The potential of the nonextremal $(Q=1)$ black hole takes a barrier-well
between the inner and
 outer horizons,  while it takes a simple barrier outside the outer
horizon. It confirmed that
the inner horizon is unstable, whereas the outer one is stable [10].
When these coalesce, a barrier-well type potential appears outside the
event horizon ($r>r_o$).
This induces the instability of the extremal black holes.
The 2D balck hole solution is symmetric in $Q_E$ and $Q_M$. The extremal
cases include three:
$Q_E= \sqrt 2, Q_M = 0; Q_E= 0, Q_M = \sqrt 2 ; Q_E= Q_M =1/\sqrt 2$. This
implies that there is
no distinction between the electrically and magnetically charged extremal
black holes [11].
It has already shown that the  2D electrically extremal  charged black hole
is unstable [5].
 This is easily recognized from our case of $Q_E= \sqrt 2, Q_M = 0$. Since
the potential
$V_{extr}(r)$ is left  invariant under the transformation
($Q_E= \sqrt 2, Q_M = 0) \to (Q_E= 0, Q_M = \sqrt 2 )$,
the magnetically charged extremal black hole is
unstable. Furthermore a symmetric combination of $Q_E= Q_M =1/\sqrt 2$ is
also unstable.

On the other hand, the Hawking temperature of a static black hole can be
calculated from (18)
\begin{equation}
T^Q_H = { \sqrt 2 \over \pi} { \sqrt{ M^2 -Q^2} \over (M + \sqrt{ M^2- Q^2})}.
\end{equation}
The Hawking temperature vanishes when $M \to Q$.

In conclusion, although the 2D  extremal charged black holes all have zero
Hawking
temperature, they  cannot be the candidates for the stable endpoint of the
Hawking evaporation.

\acknowledgments

This work was supported in part by Nondirected Research Fund, Korea
Research Foundation,1994
and by the Basic Science Research Institute Program,
Ministry of Education, Project No. BSRI-95-2413.

\figure{Fig. 1: Three graphs of generally charged black hole for $Q=0.1$
(dashed line : $-$-$-$-),
1 (dotted line  :- - - -), and $\sqrt 2$ (solid line : ---).
The corresponding event horizons are located at $r_+ = -0.004, -0.056$, and
-0.245 respectively.
When $M (= \sqrt 2) >Q (Q=0.1,1)$, the potentials outside the
event horizon are simple barriers. However  a barrier-well ($ V_{extr}(r)$)
appears
when the nonextremal
black hole (a simple barrier) approaches the extremal one($M=Q$).

 Fig.2 : The approximate  form ($V_{app}(r^*)$) of extremal black holes
outside the event horizon
($r^* = - \infty$). The aymptotically flat region is at $r^* = \infty$.
This also takes a  barrier-well type. This is localized at $r^*=0$, falls
to zero
exponentially as $r^* \to \infty$ and inverse-squarely as $r^* \to -\infty$
(solid lines).
 The dotted line is used to connect two boundaries.}
\end{document}